\documentclass[prd,aps,preprint,tightenlines,superscriptaddress,nofootinbib,showpacs,showkeys]{
 revtex4}
\usepackage{amsfonts}
\usepackage{array}
\usepackage{epsfig}
\usepackage{rotating}
\usepackage{multirow}

\usepackage{amsmath}    
\usepackage{verbatim}   
\usepackage{color}      
\usepackage{colordvi}

\usepackage{subfigure}  
\usepackage{hyperref}   

\usepackage{pstricks,rotating, graphicx}

\DeclareGraphicsExtensions{.eps,.ps}

\usepackage{pslatex}
\usepackage{txfonts}
\usepackage[latin1]{inputenc}
\usepackage[T1]{fontenc}

\newcommand{\GeV}{\ensuremath{\,\mathrm{GeV}}}

\newcommand{\msbar}{$\overline{\mathrm{MS}}\, $}

\begin{document}

\begin{titlepage}
\thispagestyle{empty}
\noindent
DESY 25--137
\hfill
October 2025 \\
\vspace{1.0cm}

\begin{center}
  {\bf \Large 
Determination of the strong coupling from high-energy data
\\
  }

  \vspace{1.25cm}
 {\large
   S.~Alekhin$^{\, a}$,
   M.V.~Garzelli$^{\, a}$,
   S.~Moch$^{\, a}$,
and  O.~Zenaiev$^{\, a}$
   \\
 }
 \vspace{1.25cm}
 {\it
   $^a$ II. Institut f\"ur Theoretische Physik, Universit\"at Hamburg \\
   Luruper Chaussee 149, D--22761 Hamburg, Germany \\
 }
  \vspace{1.4cm}
  \large {\bf Abstract}
  \vspace{-0.2cm}
\end{center}

We determine the strong coupling from high-energy data for the Drell-Yan process and top-quark hadro-production  
collected at the Large Hadron Collider and the Tevatron combined with the world data on deep-inelastic scattering (DIS) and fixed-target Drell-Yan data.
The theory description uses results at next-to-next-to-leading order in
perturbative QCD in the $\overline{\mathrm{MS}}$-scheme together with leading order QED evolution.
The DIS data are subject to stringent kinematic cuts to suppress the contribution of power corrections.
We apply a cut on the hadronic invariant mass squared $W^2 \geq 12.5~\GeV^2$ together with a series of cuts on momentum transfer squared $Q^2$.
Discarding higher-twist terms we find that the value of strong coupling $\alpha_s(m_Z)$ preferred by the data stabilizes at large enough cuts on $Q^2$, 
when low-$Q^2$ DIS data sensitive to power corrections are effectively removed.
In particular, we extract the value of $\alpha_s(m_Z,N_f=5)=0.1152 \pm 0.008$ for $N_f=5$ light flavors with the cut $Q^2>10~\GeV^2$.
In the absence of higher-twist terms less tight cuts on $Q^2$ show a clear
deterioration of the fit and lead to rising values of the strong coupling,
shifted upwards by about two standard deviations. 

\end{titlepage}
\newpage

The strong coupling $\alpha_s$, defining the strength of the interaction between quarks and gluons, 
is a fundamental parameter in quantum chromodynamics (QCD). 
The renormalized strong coupling depends on the energy scale $Q$, i.e. $\alpha_s(Q^2)$, and the prediction
of its scale dependence~\cite{Gross:1973id,Politzer:1973fx} 
dates back to the origins of QCD as the gauge theory of the strong interaction. 
The running of $\alpha_s$ is governed by the  $\beta$-function in QCD 
which is known to five loops in QCD perturbation theory~\cite{Baikov:2016tgj,Herzog:2017ohr,Luthe:2017ttg,Chetyrkin:2017bjc},
and by convention the scale $Q = m_Z$ is used for the reference value $\alpha_s(m_Z^2)$.
The determination of $\alpha_s(m_Z^2)$ from experimental data for hard-scattering processes requires the 
comparison with theory predictions, usually obtained in QCD perturbation 
theory to next-to-next-to-leading order (NNLO) in the \msbar-scheme. 
The large number of such determinations from various colliders and also 
including lattice computations is summarized by the particle data group (PDG)~\cite{ParticleDataGroup:2024cfk},
which quotes a current world average of $\alpha_s(m_Z^2) = 0.1180 \pm 0.0009$.
Knowing $\alpha_s(m_Z^2)$ as precisely as possible is a crucial task in particle physics,
e.g., to reduce parametric uncertainties in theory predictions to all processes
(cross-sections and particle decays) at the LHC and at future colliders.

The wide ranges of hard scales probed in deep-inelastic scattering (DIS) at the HERA collider 
and in fixed-target DIS experiments distinguish this process for $\alpha_s(m_Z^2)$ determinations.
However, the value of $\alpha_s(m_Z^2)$ extracted in analyses of DIS of leptons off hadrons or 
of hadron collider data from Tevatron and the Large Hadron Collider (LHC) is naturally correlated with
the description of the parton content of the proton, provided through
parton distribution functions (PDFs), as well as with the values of
heavy-quark masses $m_c$ and $m_b$ for charm and bottom quarks.
To account for such correlations global analyses are preferred based on
a combination of the world DIS data with other data sets collected at the LHC and Tevatron.
The additional hard-scattering reactions include, e.g., $W$- and $Z$-boson
production, i.e., the Drell-Yan (DY) process and top-quark hadro-production, 
which also introduces dependence on the top quark mass $m_t$.
It is therefore natural to consider simultaneous fits of the strong coupling 
$\alpha_s(m_Z^2)$, the heavy-quark masses $m_c$, $m_b$ and $m_t$ with PDFs.


In the ABMP16 analysis, this approach has been pursued in extractions of $\alpha_s(m_Z^2)$
at next-to-leading order (NLO)~\cite{Alekhin:2018pai} and at
next-to-next-to-leading order (NNLO)~\cite{Alekhin:2017kpj} in QCD,
and more recently in the ABMPtt fit at NNLO~\cite{Alekhin:2024bhs}.
A new determination of $\alpha_s(m_Z^2)$ at NNLO in QCD, combined with leading-order (LO) quantum electrodynamics (QED) evolution and including a photon PDF, will be published
shortly as a result of the ABGMZ25 PDF fit~\cite{ABGMZ25:toappear}. 
Other recent studies include work on the $\alpha_s$ and heavy-quark mass dependence
in the MSHT20 global PDF analysis~\cite{Cridge:2021qfd},
as well as determinations of $\alpha_s(m_Z^2)$ at even higher orders in perturbative QCD: specifically,
at approximate next-to-next-to-next-to-leading order (N$^3$LO) in global PDF fits by MSHT~\cite{Cridge:2024exf}
and by NNPDF~\cite{Ball:2025xgq}, in the latter case combined with NLO QED evolution.
Some of these determinations exhibit tensions among each other within the
quoted uncertainties and with the world average provided by the PDG, 
underpinning the importance of detailed studies to achieve a consolidated view. 

The differences in the extracted values of $\alpha_s$ can be attributed both
to the theoretical treatment of the hard scattering process beyond standard
QCD factorization at leading twist, and to the selection of input data sets.
In particular, the inclusion or exclusion of DIS data at low values of the space-like momentum $Q^2$ 
transferred by the virtual boson plays a significant role.
At large $Q^2$ DIS can be described in the operator product expansion (OPE) within the leading-twist 
approximation, but at scales comparable to the proton mass $m_P$ additional power corrections of order $1/Q^2$ become important.
These include target mass corrections (TMCs), which are purely kinematic, and higher-twist (HT) terms, which originate from genuine multi-parton correlations inside the proton.
TMCs arising from kinematic effects of the finite proton mass $m_P$ can be computed in a model-independent way using the OPE.
The Georgi--Politzer prescription~\cite{Georgi:1976ve} provides explicit
formulas for the suitably corrected DIS structure functions $F_1$ and $F_2$,
expressed in terms of the Nachtmann variable~\cite{Nachtmann:1973mr} $\xi = 2x/(1+\gamma)$ with $\gamma = \sqrt{1+4x^2 m_P^2/Q^2}$ 
and $x$ the standard Bjorken variable. 
TMCs thus modify the scaling behavior of structure functions without introducing new dynamical information.
HT contributions can significantly affect the structure functions $F_1$ and $F_2$ at low $Q^2$ and large $x$.

Analogous to leading-twist terms, HT contributions can also be written as convolutions of process-dependent coefficient functions with process-independent HT partonic operator matrix elements (OMEs). 
The HT OMEs are derived from local composite operators that involve more external quark and gluon fields than twist-two operators.
Since multiple OMEs can contribute at the same twist level, the resulting convolutions involve sums over operator indices as well as integrals over several momentum fractions. 
Most of these momentum fractions are not directly accessible experimentally. 
Consequently, the dependence of HT OMEs on these momentum fractions cannot be fitted, leading to a more complex situation than in the case of (leading-twist) PDFs.
Furthermore, HT OMEs and their corresponding Wilson coefficients satisfy different renormalization group equations, so their scale evolution must be handled on a case-by-case basis and separately from the twist-two contributions.
As a result, HT contributions are added linearly -- not multiplicatively, cf. e.g.~\cite{Harland-Lang:2025wvm} -- to the leading twist terms in the structure functions.
In particular, twist-four contributions to DIS, known since many years~\cite{Jaffe:1982pm,Ellis:1982cd}
together with a framework for their evolution~\cite{Bukhvostov:1985rn,Braun:2009vc}, are relevant.
However, due to the large number of HT OMEs, their particular momentum fraction dependence and the limited availability of experimental data, 
their shapes remain poorly constrained.
Therefore, they are treated  phenomenologically in global fits.  
For this reason, the ABMP16, ABMPtt and ABGMZ25 PDF analyses of DIS data at low $Q^2$ (and earlier variants~\cite{Alekhin:2012ig}), have taken the
practical approach to write the DIS structure functions for $i=1,2$ as
\begin{equation}
\label{eq:dis-sf+ht}
  F_i(x,Q^2) = F^{\text{TMC}}_i(x,Q^2) 
    + \frac{H^{\tau=4}_i(x)}{Q^2} 
    + \frac{H^{\tau=6}_i(x)}{Q^4} + \dots\, ,
\end{equation}
where $F^{\text{TMC}}_i$ are the leading-twist TMC corrected structure functions and the additive HT terms $H_i(x)$ are fitted functions of $x$, constrained to vanish at $x=1$.
Details are described in~\cite{Alekhin:2017kpj,Alekhin:2012ig}, and previously, specifically in the context of the determination of the strange sea distribution from charged-current neutrino-nucleon DIS, in~\cite{Alekhin:2008mb,Alekhin:2008ua}.
In practice, only twist-four terms are retained, since twist-six contributions
are negligible in the kinematic range of DIS data considered, see e.g.~\cite{Alekhin:2007fh}.

In this letter, we explore strategies to reduce the impact of power corrections on the analysis of low-$Q^2$ DIS data and the determination of $\alpha_s$.
To that end we perform an NNLO QCD~+~LO QED analysis of DY and top-quark data
collected at the LHC and Tevatron, combined with world DIS data and fixed-target DY data, following the approach of the ABGMZ25 PDF fit~\cite{ABGMZ25:toappear}. 
To suppress contributions from power corrections, we impose a series of
stringent cuts on the DIS kinematics, specifically on the space-like momentum $Q^2$  transferred by the virtual boson and the invariant mass of the hadronic system
\begin{eqnarray}
  \label{eq:wdef}
  W^2 = m_P^2 + Q^2 (1-x)/x
  \, ,
\end{eqnarray}
that might eliminate the HT terms.
The resulting impact on the extracted value of $\alpha_s(m_Z^2)$ is systematically quantified.

\begin{figure}[t!]
\centering 
\includegraphics[width=\textwidth]{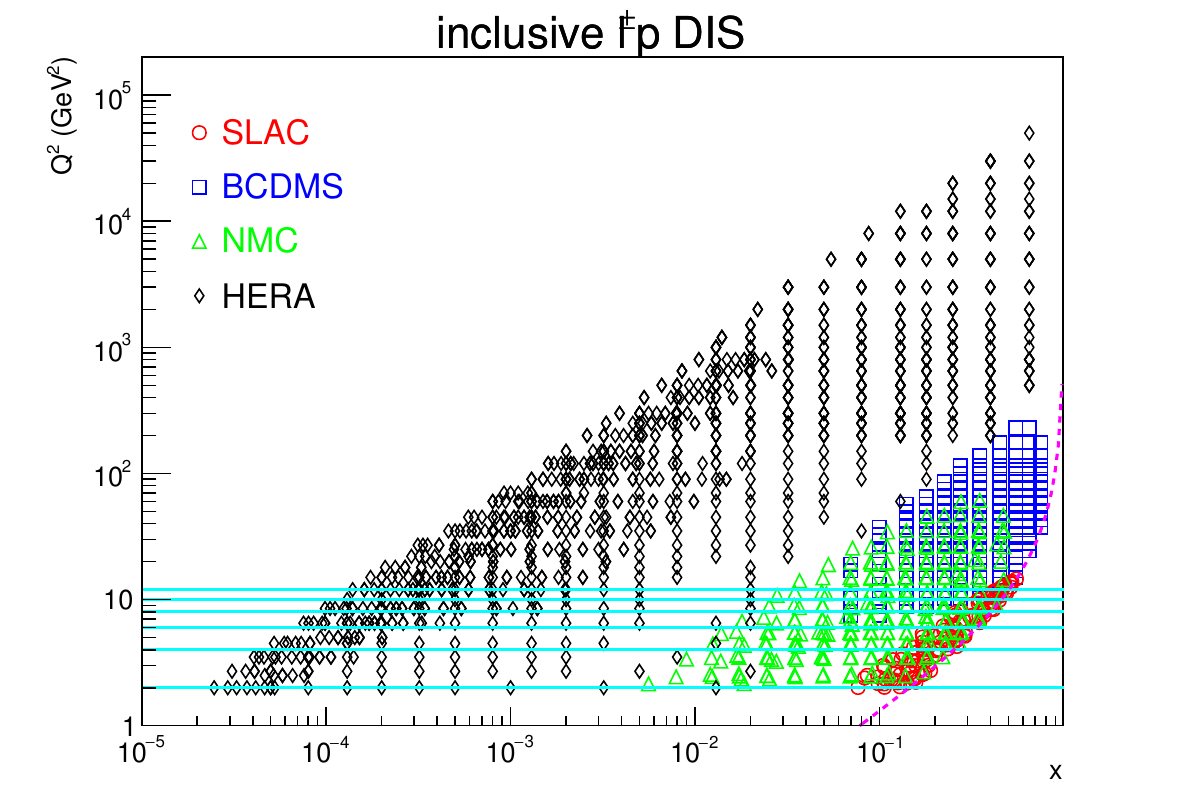}
\hfill
\caption{\label{fig:kin} 
  Kinematics of the inclusive DIS data used in the present analysis in
  terms of $Q^2$ and Bjorken $x$ (circles: SLAC~\cite{Whitlow:1990gk},
  squares: BCDMS~\cite{BCDMS:1989eab}, triangles: NMC~\cite{NewMuon:1996fwh},
  diamonds: HERA~\cite{H1:2015ubc}).
  The turquoise horizontal solid lines display the $Q^2$ cuts considered in the present analysis.
  The data points plotted at large-$x$ and low $Q^2$ all fulfill the
  constraint  $W^2 \geq 12.5 \GeV^2$ indicated by the purple dashed line.
}
\end{figure}
The theoretical treatment in the ABGMZ25 fit uses NNLO QCD predictions as described
in~\cite{Alekhin:2017kpj} together with LO QED evolution~\cite{ABGMZ25:toappear}.
In particular, it employs the fixed-flavor-number scheme (FFNS)
with a fixed number $N_f$ of light-quark flavors and massive heavy-quark DIS structure functions included at 
(approximate) NNLO as described in~\cite{Alekhin:2017kpj,Kawamura:2012cr}
(see \cite{Ablinger:2024xtt} and references therein for the latest developments).
The value of the strong coupling is then determined as $\alpha_s(m_Z,N_f=5)$
with $N_f=5$ light-quark flavors in the FFNS, while 
$\alpha_s(m_Z,N_f)$ for other values of $N_f$ is obtained by the standard decoupling, see. e.g.\ \cite{Herren:2017osy}.
The world charged-lepton DIS data considered in the ABGMZ25 PDF fit has been taken at the  
SLAC~\cite{Whitlow:1990gk}, BCDMS~\cite{BCDMS:1989eab},
NMC~\cite{NewMuon:1996fwh} and HERA~\cite{H1:2015ubc} experiments.
Specifically, data collected on proton targets are used to avoid nuclear corrections (such as those required for deuterium targets).
The separation of light-flavor PDFs is achieved instead with the help of  DY data from the LHC through precise measurements of $W$ and $Z$ boson production.
The kinematic coverage of the charged-lepton DIS data in terms of $Q^2$ and Bjorken $x$ is shown in Fig.~\ref{fig:kin}.
The DIS collider data taking at HERA was achieved at an unprecedented precision of $O(1\%)$,
while the normalization of earlier fixed-target DIS experiments has been treated as a free parameter.
The fitted values of their normalization factors are determined with an uncertainty of $O(1\%)$,
and details have been discussed in~\cite{Alekhin:2012ig,Alekhin:2017kpj}.

We perform variants on the procedure adopted for the ABGMZ25 fit with cuts $Q^2 \geq Q_{\rm min}^2$ 
choosing the series $Q_{\rm min}^2 \in \{2, 4, 6, 8, 10, 12\}~\GeV^2$
together with a stringent cut on the invariant mass in Eq.~(\ref{eq:wdef}), $W^2 = 12.5 \GeV^2$,
and setting the HT terms $H_i(x)$ in Eq.~(\ref{eq:dis-sf+ht}) to zero in the description of all charged-lepton DIS data.~\footnote{
It should be noted that the  $Q^2$ cut is applied only to the charged-lepton DIS data sets and not to neutrino-nucleon DIS data.
While the neutrino data have little impact on the determination of $\alpha_s$, they play an important role in constraining the strangeness PDFs.
The charged-current DIS description of neutrino-nucleon data includes 
in the fit the HT terms determined in \cite{Alekhin:2008mb,Alekhin:2008ua,Alekhin:2007fh}.}
Cuts on the invariant mass of the hadronic system $W^2$ in the range $W^2 = 12.5 \div 15 \GeV^2$ 
and on $Q_{\rm min}^2 \gtrsim 10~\GeV^2$ have long been proposed, see, e.g., \cite{Martin:2002dr,Blumlein:2006be},
and were previously applied in variants of the ABMP16 PDF fit~\cite{Alekhin:2017kpj} (see also \cite{Alekhin:2012ig}).
The $Q_{\rm min}^2$ cuts are indicated by horizontal lines in
Fig.~\ref{fig:kin} and they gradually remove most of the SLAC data at low $Q^2$ and large $x \gtrsim 0.1$, 
parts of the NMC data in the range $10^{-2} \lesssim x \lesssim 0.1$, as well as
the HERA data down to $x$-values $x \gtrsim 10^{-5}$.

Besides world DIS data, the series of PDF fits (ABMP16, ABMPtt, and ABGMZ25)
utilize an extensive collection of high-precision experimental data. 
This includes DY measurements from both collider experiments (Tevatron and
LHC) and fixed-target, non-collider experiments.
The fixed-target DY data, in particular, provide valuable constraints on the sea quark
distributions at high $x$, and allow for a better separation of the $\bar u$- and $\bar d$-quark components.
In addition, the fit also incorporates inclusive cross-section data for top-quark pair
production and single-top production from Tevatron and LHC.~\footnote{The DY data sets are listed in Tabs.~II, III and V
  in~\cite{Alekhin:2017kpj} and the data on the inclusive cross-section 
  for production of top-quark pairs and single top-quarks are given in Figs.~20 and 21 of~\cite{Alekhin:2024bhs}.
}
PDFs are parametrized at the starting scale with flexible functional forms, including parameters for both small- and large-$x$ behavior and polynomial terms for full $x$-range adaptability.
These serve as boundary conditions for the QCD evolution, cf.\ e.g.~\cite{Alekhin:2017kpj}.
The quality of the fit is carefully checked by introducing additional terms in the parametrization at the starting scale as needed to further improve the data description.
This procedure has led to continuous refinements of the PDF fits (ABMP16, ABMPtt, and ABGMZ25), demonstrating that the chosen approach is adequate for the analysis.

\begin{figure}[ht]
\centering 
\includegraphics[width=\textwidth]{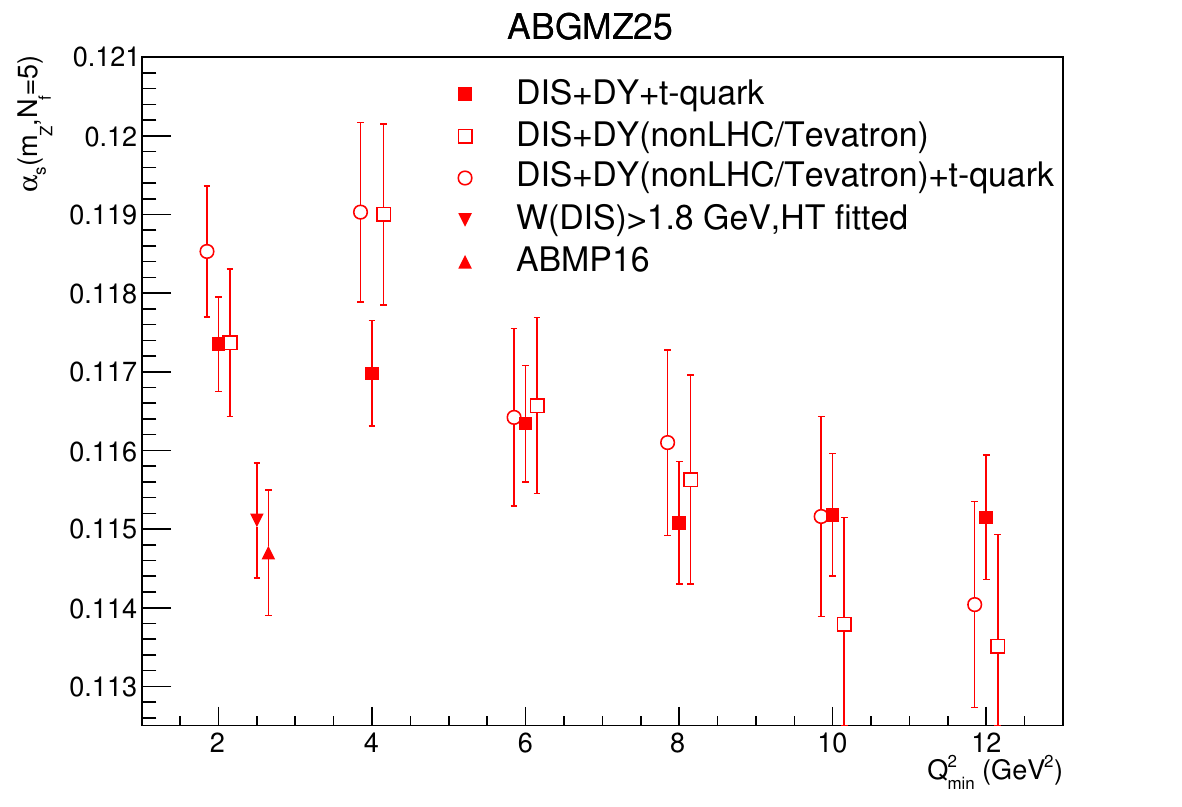}
\hfill
\caption{\label{fig:alp} 
  The value of $\alpha_s(m_Z,N_f=5)$ obtained in variants of present
  analysis based on different combinations of data
  (solid squares: DIS+DY+$t$-quark data;
  hollow squares: the same with the high-energy data from the LHC and Tevatron dropped;
  hollow circles: DIS, DY fixed-target and $t$-quark data),
  with the cuts of $W^2>12.5~\GeV^2$ and $Q^2>Q_{\rm min}^2$ imposed on the DIS data and the HT set to zero.
  For comparison the values of $\alpha_s(m_Z,N_f=5)$ obtained in the ABMP16 fit~\cite{Alekhin:2017kpj}
  (upward triangle) and in the present analysis with the cuts
  $W>1.8~\GeV$, $Q^2>2.5~\GeV^2$ and the HT terms
  modeled as in Ref.~\cite{Alekhin:2017kpj} are given (downward triangle). 
}
\end{figure}
The value of the strong coupling, 
with the increasing $Q_{\rm min}^2$-cuts are displayed in Fig.~\ref{fig:alp}.
Several variants of the present analysis, each based on different combinations of input data, are shown.
Results with solid squares denote fits based on the complete set of DIS, DY, and top-quark data.
There is a clear trend of decreasing values of $\alpha_s(m_Z)$ as the $Q_{\rm min}^2$-cut increases,
recall we also use the cut $W^2 = 12.5 \GeV^2$ throughout.
We obtain the values $\alpha_s(m_Z,N_f=5,Q_{\rm min}^2=2\GeV^2)=0.1173  \pm 0.0007$ and 
$\alpha_s(m_Z,N_f=5,Q_{\rm min}^2=10\GeV^2)=0.1152  \pm 0.0008$, which are in tension by about $2\sigma$.
For cuts $Q_{\rm min}^2 \geq 10 \GeV^2$ the extracted values of $\alpha_s(m_Z,N_f=5)$ stabilize in Fig.~\ref{fig:alp}.
This is a clear indication of the importance of accounting for HT terms in fits
of low-$Q^2$ DIS data, recall they are set to zero here.
It is also a confirmation of the effectiveness of the cut $Q_{\rm min}^2 \geq 10 \GeV^2$ previously applied to remove data sensitive to HT effects.
Other variants considered in Fig.~\ref{fig:alp} represent the same fit, but with high-energy (LHC
and Tevatron collider) data removed (hollow squares)
as well as fits in which the top-quark data are included together with the DIS and fixed-target DY ones (hollow circles).
They all show a largely consistent picture, with a slightly increased uncertainty in the extracted values of $\alpha_s(m_Z)$ 
since the number of data points (NDP) decreases, when selected sets of data are dropped.
Some small fluctuations of those fits at the level of $1\sigma$ are visible,
in particular for $Q_{\rm min}^2 = 4\GeV^2$, which are due to the data quality.
For comparison, the values from a fit under the same conditions but using looser cuts ($W > 1.8~\GeV$, $Q_{\rm min}^2 = 2.5~\GeV^2$)
and modeling the HT terms as in Ref.~\cite{Alekhin:2017kpj} (downward triangle)
and the one of the ABMP16 PDF fit~\cite{Alekhin:2017kpj} (upward triangle) are also shown.
This fit shows consistent results with those imposing tight cuts $W^2 = 12.5 \GeV^2$ and $Q_{\rm min}^2 \geq 10 \GeV^2$.
Thus, the use of low-$Q^2$ DIS data in global PDF fits either requires modeling of HT terms of twist-four in the fits,
or, alternatively removal of those data through tight cuts.
As a result, we choose the value of $Q_{\rm min}^2 = 10 \GeV^2$ as the default
in our extraction of $\alpha_s(m_Z)$.
With such a stringent $Q_{\rm min}^2 = 10 \GeV^2$ cut, the influence of power corrections on the DIS data is minimized, while the
inclusion of high-energy collider data helps to constrain the PDFs,
leading to a more precise determination of $\alpha_s(m_Z)$.

\begin{table}[t!]
  \begin{center}
  \renewcommand{\arraystretch}{1.3}  
  \begin{tabular}{|c|r|r|r|r|r|}
    \hline
$Q_{\rm min}^2~(\GeV^2)$ & \multicolumn{4}{c|}{$\chi^2/NDP$} \\ 
    \hline    
  &  \multicolumn{1}{|c|}{SLAC~\cite{Whitlow:1990gk}}& \multicolumn{1}{|c|}{BCDMS~\cite{BCDMS:1989eab}}& \multicolumn{1}{|c|}{NMC~\cite{NewMuon:1996fwh}}& \multicolumn{1}{|c|}{HERA~\cite{H1:2015ubc}}\\
\hline
2 &$230/124= 1.85$ & $369/337= 1.09$ &$441/251= 1.76$ &$1569/1185=1.32$ \\
\hline
4 &$162/79= 2.05$ &$367/337= 1.09$ &$326/201= 1.62$ &$1447/1120=1.29$ \\
\hline
6 &$92/50= 1.84$ &$359/337= 1.07$&$241/148=1.63$ &$1361/1092=1.25$ \\
\hline
8 &$52/34= 1.53$ &$353/335= 1.05$&$168/118=1.42$&$1297/1056=1.23$ \\
\hline
{\bf 10}&${\bf 24/22= 1.09}$ &${\bf 354/330= 1.07}$ &${\bf 120/92= 1.30}$ &${\bf 1234/1016= 1.21}$ \\
\hline
12&$3/10= 0.30$ &$324/313= 1.04$&$97/70= 1.39$&$1221/1007= 1.21$ \\
\hline
  \end{tabular}
\caption{\small 
\label{tab:datahq}
      {The values of $\chi^2/NDP$ obtained for the inclusive DIS data sets
        used in variants of present analysis with various cuts on $Q^2$.
        The row with bold-faced fonts displays our nominal fit.
}
}
\end{center}
\end{table}
In Tab.~\ref{tab:datahq} we summarize the values of $\chi^2/NDP$ obtained for the inclusive charged-lepton DIS data sets
used in variants of the present analysis with various cuts on $Q^2$.
These partial $\chi^2/NDP$ per data set correspond to the $\alpha_s(m_Z)$ values plotted with solid squares in Fig.~\ref{fig:alp}.
The row in bold in Tab.~\ref{tab:datahq} indicates our nominal fit.
The $\chi^2/NDP$ values in Tab.~\ref{tab:datahq} especially for the SLAC and
NMC data sets and the low $Q_{\rm min}^2$-cuts also indicate tension due to
inadequate theory description (missing HT terms).
A qualitatively similar trend in the $Q^2$-cut dependence of the values of $\alpha_s(m_Z)$
in Fig.~\ref{fig:alp} has recently been observed in an update to the MSHT global PDF fit~\cite{Harland-Lang:2025wvm},
which focuses on the high-$x$ region.
The inclusion of TMC and HT corrections at NNLO and approximate N$^3$LO was found to have a moderate but non-negligible impact on both the PDFs
and the extracted strong coupling, although the details differ with respect to our analysis due to the different theoretical modeling used
(e.g. multiplicative HT terms in~\cite{Harland-Lang:2025wvm}).

The complete set of Standard Model (SM) parameters obtained at NNLO in QCD together with LO QED evolution
in the nominal analysis with $Q_{\rm min}^2 = 10~\GeV^2$
(cf. the $\alpha_s(m_Z)$ value plotted with solid squares in Fig.~\ref{fig:alp})
and HT terms set to zero are the following ones: 
\begin{eqnarray}
\label{eq:res-new}
  \alpha_s(m_Z,N_f=5)&=&0.1152 \pm 0.0008\, ,\nonumber \\
  m_t(m_t)&=&161.1 \pm 0.7\phantom{000} \GeV \, ,\nonumber\\
  m_b(m_b)&=&3.94\phantom{0} \pm 0.10\phantom{00} \GeV \, ,\nonumber\\
  m_c(m_c)&=&1.256 \pm 0.019\phantom{0} \GeV \, .
\end{eqnarray}
As in previous fits, the heavy-quark masses $m_c$ and $m_b$ are mostly
constrained by the semi-inclusive HERA data on DIS charm-~\cite{H1:2012xnw} and bottom-quark~\cite{H1:2009uwa,ZEUS:2014wft} production, see also~\cite{Zenaiev:2015qea}. 
For comparison, the values of SM parameters obtained in the variant 
with looser cuts of $W > 1.8~\GeV$ and $Q_{\rm min}^2 =  2.5~\GeV^2$, along with a
simultaneous fit of twist-four terms
(cf. the $\alpha_s(m_Z)$ value plotted with a downward triangle in Fig.~\ref{fig:alp}) 
are as follows:
\begin{eqnarray}
\label{eq:res-ht-fitted}
\alpha_s(m_Z,N_f=5) &=& 0.1151  \pm 0.0007\, ,\nonumber \\
m_t(m_t) &=& 161.7 \pm 0.8\phantom{000} \GeV \, ,\nonumber\\
m_b(m_b) &=& 3.94\phantom{0} \pm 0.10\phantom{00} \GeV \, ,\nonumber\\
m_c(m_c) &=& 1.254 \pm 0.018\phantom{0} \GeV \, .
\end{eqnarray}
The previous ABMP16 PDF fit, using the same methodology (loose cuts $W > 1.8~\GeV$, $Q > 2.5~\GeV^2$ and the HT terms fitted) 
quotes values well compatible with Eq.~(\ref{eq:res-ht-fitted}):
\begin{eqnarray}
\label{eq:res-abmp16}
\alpha_s(m_Z,N_f=5) &=&  0.1147 \pm 0.0008\, ,\nonumber \\
m_t(m_t) &=& 160.9 \pm 1.1\phantom{000} \GeV \, ,\nonumber\\
m_b(m_b) &=& 3.84\phantom{0} \pm 0.12\phantom{00} \GeV \, ,\nonumber\\
m_c(m_c) &=& 1.252 \pm 0.018\phantom{0} \GeV \, .
\end{eqnarray}
The recent update with the ABMPtt PDFs at NNLO in QCD
using double-differential data for top-quark pair production
also gave consistent results
\begin{eqnarray}
  \label{eq:res-abmptt}
\alpha_s(m_Z,N_f=5) &=&  0.1150 \pm 0.0009\, ,\nonumber \\
m_t(m_t) &=& 160.6 \pm 0.6\phantom{000} \GeV \, ,
\end{eqnarray}
and values for $m_c(m_c)$ and $m_b(m_b)$ very similar to those obtained in
the ABMP16 fit, cf. Eq.~(\ref{eq:res-abmp16}).

In summary, the inclusion of high-energy collider data from the LHC and Tevatron in a combined fit of PDFs, heavy-quark masses and strong coupling has enabled a new approach to study and quantify the impact of power corrections, which are inherent to the theoretical description of low-$Q^2$ DIS data.
By applying a set of tight cuts on the minimum virtuality of the gauge boson as well as on the invariant mass of the hadronic system, data with significant sensitivity to HT effects are excluded from the analysis. 
The use of high-energy collider data -- particularly DY rapidity distributions extending to forward rapidities and thus large-$x$ values -- helps to compensate for the reduced statistics that result from these cuts.
The extracted value of $\alpha_s(m_Z)$ obtained in a FFNS at NNLO
is consistent with a fit that applied looser cuts -- including low-$Q^2$ data and fitting twist-four terms simultaneously -- and earlier results.
This approach settles a long-standing open problem in $\alpha_s(m_Z)$ extractions, paving the way for improved consistency among the different determinations published by the community and for their inclusion in the PDG review.
Future work will be devoted to repeating this analysis at N$^3$LO accuracy.

\acknowledgments
\noindent
M.V.G. thanks the participants of the PDF4LHC with PHYSTAT 2025 Workshop~\footnote{\url{https://indico.cern.ch/event/1553776/}} for interesting discussions.
The work of M.V.G. and S.M. has been supported in part by the Deutsche Forschungsgemeinschaft through the Research Unit FOR 2926, {\it Next  Generation pQCD for Hadron Structure: Preparing for the EIC}, project number 40824754.
The work of O.~Z. has been supported by the {\it MSCA4Ukraine Programme} of the European Commission through the Alexander von Humboldt foundation.


\end{document}